\documentclass[prd,preprint,superscriptaddress,showpacs,amsmath,amssymb]{revtex4-1}

\begin{document}

\title {Stable Bimetric Theories}
 
\author{Idan Talshir}
\email[E-mail:]{talshir@post.tau.ac.il}

\affiliation{School of Physics, Tel Aviv University, Ramat Aviv 69978, Israel}

\begin{abstract}
We prove that there are energetically stable bimetric theories. These theories satisfies a positive energy theorem. We construct a model example.
\end{abstract}
\maketitle

\section{Introduction}

Bimetic theories are gravity theories with two second rank tensor dynamical fields. These tensors may be used to construct affine connections. The free action for each tensor is the Einstein-Hilbert action. Some of these theories have been proposed as a modification of general relativity designed to predict cosmological and astrophysical phenomena.\\

It has been proven that a large set of multiple-metric theories that can be approximated by the Pauli-Fierz action suffer from Boulware-Deser ghost instabilities \cite{inconsistencies}. A sub-set of these theories that avoid the ghost problem has been constructed \cite{RandH}, where the structure of the interaction yields a constraint that removes the ghost. However, recent developments \cite{SandA} cast doubt on the relevance of these theories, because the same constraint gives rise to superluminal shock waves. It has also been proved \cite{BMV} that in this specific interaction the null energy condition is violated. The null energy condition is  necessary for the dominant energy condition, which has a significant role in the energy stability of the theory, as we demonstrate in the following.\\
We want to construct a model for which: 1) The total canonical energy for all the solutions with asymptotic boundary conditions is non-negative. and 2) The total energy is zero only in the vacuum state, i.e., for which both metrics are Minkowskian and other fields have zero amplitude. First we review results from a previous paper \cite{Talshir} about the positive energy theorem for multi-metric theories.\\
\paragraph{}The general action for a bimetric theory can be written in the form
 \begin{equation} 
 \label{one}
I = \int {{\cal L}{d^4}x = } \int {(\alpha {{\cal L}_G} + \beta {{\hat {\cal L}}_G} + {{\cal L}_I}){d^4}x} 
 \end{equation}
 
 where the scalar densities in the parentheses are the Einstein-Hilbert scalar density for the metric  
 ${g_{\mu \nu }}$,${{\cal L}_G} =  - \frac{1}{{16\pi G}}R\sqrt {  g} $, a scalar density for the twin metric ${\hat g_{\mu \nu }}$, ${{\hat {\cal L}}_G} =  - \frac{1}{{16\pi G}}\hat R\sqrt {  \hat g} $ ($g,\hat{g}$ are the negatives of the determinants of the metric and twin metric respectively), and an interaction term which depends on both metric and other fields and their derivatives, i.e., $ {{\cal L}_I}[{g_{\mu \nu }};{g_{\mu \nu ,\rho }};{{\hat g}_{\mu \nu }};{g_{\mu \nu ,\rho }};\psi ;{\psi _{,\rho }}]$, where $\psi $ symbolizes any other fields with any tensorial properties. The coefficients $\alpha $ and $\beta$ are any constants. We notice that the derivatives of the metric (and other fields) are not tensors, but they can be arranged to construct a tensor, e.g. the difference between the Christoffel connections of the to metrics, and to construct a scalar density ${{\mathcal{L}}_{I}}$.   \\
 The canonical energy is the space integration of the time-time component of the canonical energy momentum pseodo-tensor $\Theta _\beta ^\alpha$ which is the conserved current we get from invariance of the action with respect to coordinate translations.
 \begin{equation} \label{two}
\Theta _\beta ^\alpha  =  - L\delta _\beta ^\alpha  + {g_{lm,\alpha }}\frac{{\partial L}}{{\partial {g_{lm,\beta }}}} + {{\hat g}_{lm,\alpha }}\frac{{\partial L}}{{\partial {{\hat g}_{lm,\beta }}}} + {\psi _{,\alpha }}\frac{{\partial L}}{{\partial {\psi _{,\beta }}}}
\end{equation}
It has been shown \cite{Talshir} that with an appropriate asymptotic conditions on the fields and derivative of the Lagrangian with respect to field derivatives, the total energy is the sum of canonical energy expressions for each metric separately.
Also, assuming asymptotic boundary conditions for the interaction energy momentum tensors
\[T_\mu ^\nu  = {\rm O}({r^{ - 4}})\,\,\,\,\,\,\,\,\,\,\hat T_\mu ^\nu  = {\rm O}({r^{ - 4}})\]
and for the metric fields
\[{g_{\mu \nu }} \simeq {\eta _{\mu \nu }} + {\rm O}({r^{ - 1}})\,\,\,\,,\,\,\,\,{\hat g_{\mu \nu }} \simeq {\eta _{\mu \nu }} + {\rm O}({r^{ - 1}})\]
then the total energy momentum vector is a sum of two Arnowitt-Deser-Misner (ADM) expressions, surface integrals of metric derivatives:
 \begin{equation} \label{twothreeone}
P_{total}^\nu  = {P^\nu } + {{\hat P}^\nu }
\end{equation}
where
\begin{equation} \label{twothreetwo}
{P^0} \equiv \frac{\alpha }{{16\pi G}}\int {\left( {\frac{{\partial {g_{ij}}}}{{\partial {x^j}}} - \frac{{\partial {g_{jj}}}}{{\partial {x^i}}}} \right)} d{s^i}{\mkern 1mu} {\mkern 1mu} {\mkern 1mu} {\mkern 1mu} {\mkern 1mu} {\mkern 1mu} {\mkern 1mu} {\mkern 1mu} {\mkern 1mu} {\mkern 1mu} {{\hat P}^0} \equiv \frac{\beta }{{16\pi G}}\int {\left( {\frac{{\partial {{\hat g}_{ij}}}}{{\partial {x^j}}} - \frac{{\partial {{\hat g}_{jj}}}}{{\partial {x^i}}}} \right)} d{s^i}
\end{equation}
and
\begin{equation} \label{twothreethree}
{P^j} \equiv \frac{\alpha }{{16\pi G}}\int {\left( {\frac{{\partial {g_{kk}}}}{{\partial {x^0}}}\frac{{\partial {g_{k0}}}}{{\partial {x^k}}}{\delta _{ij}} + \frac{{\partial {g_{j0}}}}{{\partial {x^i}}} - \frac{{\partial {g_{ij}}}}{{\partial {x^0}}}} \right)d{s^i}{\mkern 1mu} {\mkern 1mu} {\mkern 1mu} {\mkern 1mu} {\mkern 1mu} {\mkern 1mu} {\mkern 1mu} {\mkern 1mu} } {{\hat P}^j} \equiv \frac{\beta }{{16\pi G}}\int {\left( {\frac{{\partial {{\hat g}_{kk}}}}{{\partial {x^0}}}\frac{{\partial {{\hat g}_{k0}}}}{{\partial {x^k}}}{\delta _{ij}} + \frac{{\partial {{\hat g}_{j0}}}}{{\partial {x^i}}} - \frac{{\partial {{\hat g}_{ij}}}}{{\partial {x^0}}}} \right)d{s^i}{\mkern 1mu} {\mkern 1mu} {\mkern 1mu} {\mkern 1mu} {\mkern 1mu} {\mkern 1mu} {\mkern 1mu} {\mkern 1mu} } 
\end{equation}
The total energy momentum vector $P^\mu$ is Lorentz covariant, i.e., a vector under
Lorentz transformations of the coordinate system. It is also invariant under any coordinate transformation that tends to the identity at infinity, and preserves the boundary conditions.\\
Based on these results, Witten's proof \cite{Witten},\cite{Faddeev} of the positive energy theorem for general relativity has been generalized for bimetric systems \cite{Talshir} and it has been shown that if the coefficients $\alpha,\beta$ are positive and the energy momentum tensors ${{T}_{\mu \nu }},{{{\hat{T}}}_{\mu \nu }}$ obey the dominant energy condition then the total canonical energy is not negative, and has zero value only for a ground state. A ground state is obtained when all metrics are Minkowskian and all energy momentum tensors are zero.\\
 
\section{Construction of a positive energy bimetric Lagrangian}

In order to construct a model that maintains the bimetric positive energy theorem conditions, we demand that the energy momentum tensor for each metric separately obeys the dominant energy condition. In the proof of the positive energy theorem for general relativity it is \emph{assumed} that the energy momentum tensor obeys the dominant energy condition, which is a correct assumption for standard matter fields. 
In the bimetric model the energy momentum tensor for each metric is (also) derived from a coupling with the other metric, and in general one can not assume that the dominant energy condition is maintained, because the energy momentum tensor depends on the configuration of the specific metrics. Here we give an example of an interaction for which the derived energy momentum tensors obey the dominant energy condition by their own structure, regardless of the specific state. That special demand can be achieved if the energy momentum tensors are in the form: 
\begin{equation} 
\begin{gathered}{T_{\mu\nu}}=-F[{g_{\mu\nu}},{{\hat{g}}_{\mu\nu}},{g_{\mu\nu,\rho}},{{\hat{g}}_{\mu\nu,\rho}},\psi,{\psi_{,\rho}}]{g_{\mu\nu}}\hfill\\
{{\hat{T}}_{\mu\nu}}=-\hat{F}[{g_{\mu\nu}},{{\hat{g}}_{\mu\nu}},{g_{\mu\nu,\rho}},{{\hat{g}}_{\mu\nu,\rho}},\psi,{\psi_{,\rho}}]{{\hat{g}}_{\mu\nu}}\hfill
\end{gathered}
\end{equation}
The scalars $F,\hat{F}$ are functionals of the various fields. We demand that these scalars are not negative for every field configuration:
\begin{equation}
F\ge 0\,\,\,,\,\,\,\hat{F}\ge 0
\end{equation}
 In this way the dominant energy condition is automatically maintained for both energy momentum tensors. This is because by definition the dominant energy condition is maintained when ${{T}_{\mu \nu }}{{u}^{\mu }}{{v}^{\nu }}\ge 0,{{{\hat{T}}}_{\mu \nu }}{{{\hat{u}}}^{\mu }}{{{\hat{v}}}^{\nu }}\ge 0$ for every time-like future-pointing vectors ${{u}^{\mu }},{{v}^{\nu }}$ and ${{\hat{u}}^{\mu }},{{\hat{v}}^{\nu }}$, and when ${{T}^{\mu \nu }}{{u}_{\nu }},{{\hat{T}}^{\mu \nu }}{{\hat{u}}_{\nu }}$ are time-like vectors for every pair of time-like vectors ${{u}_{\nu }},{{{\hat{u}}}_{\nu }}$. One should keep in mind that the definition of a vector as time-like, is, as for any causal character, with respect to the corresponding metric, so if ${{u}_{\nu }},{{{\hat{u}}}_{\nu }}$ are time-like with respect to the metrics ${{g}_{\mu \nu }},{{{\hat{g}}}_{\mu \nu }}$ respectively, then ${{T}^{\mu \nu }}{{u}_{\nu }}=(-F){{g}^{\mu \nu }}{{u}_{\nu }},{{{\hat{T}}}^{\mu \nu }}{{{\hat{u}}}_{\nu }}=(-F){{{\hat{g}}}^{\mu \nu }}{{{\hat{u}}}_{\nu }}$ are time-like vectors. In addition, two time-like vectors ${{u}^{\mu }},{{v}^{\nu }}$ belong to the same equivalence class of future-pointing or past-pointing if ${{g}_{\mu \nu }}{{u}^{\mu }}{{v}^{\nu }}\le 0$ (and respectively for ${{{\hat{u}}}^{\mu }},{{{\hat{v}}}^{\nu }}$) when ${{{\hat{g}}}_{\mu \nu }}{{{\hat{u}}}^{\mu }}{{{\hat{v}}}^{\nu }}\le 0$, so that
\begin{equation}
\begin{gathered}{T_{\mu\nu}}{u^{\mu}}{v^{\nu}}=-F{g_{\mu\nu}}{u^{\mu}}{v^{\nu}}\geqslant0\hfill\\
{{\hat{T}}_{\mu\nu}}{{\hat{u}}^{\mu}}{{\hat{v}}^{\nu}}=-\hat{F}{{\hat{g}}_{\mu\nu}}{{\hat{u}}^{\mu}}{{\hat{v}}^{\nu}}\geqslant0\hfill
\end{gathered}
\label{gthree}
\end{equation}
A last condition is that for the energy momentum tensors to be consistent with the boundary conditions we have to assume on the metrics, in order to define the total canonical energy, the condition
\begin{equation}
\label{gtwo}
T{{_{\nu }^{\mu }}_{r\to \infty }}=\hat{T}{{_{\nu }^{\mu }}_{r\to \infty }}=O({{r}^{-4}})
\end{equation}
so that
\begin{equation}
\label{efive}
{{F}_{r\to \infty }}={{{\hat{F}}}_{r\to \infty }}=O({{r}^{-4}})
\end{equation}
This condition can be achieved for, for example, by constructing a scalar from the metrics and requiring that
 \[F\left( {{g}^{\mu \nu }}\to {{\eta }^{\mu \nu }},{{{\hat{g}}}^{\mu \nu }}\to {{\eta }^{\mu \nu }} \right)=O({{(g-1)}^{4}})\]
and the same for ${\hat{F}}$, or by a coupling to another field which has zero amplitude on the boundary. In the following we choose the second option.\\
For simplicity, we choose to construct an interaction without explicit dependence on metric derivatives:
\begin{equation}
\label{fone}
{{T}_{\mu \nu }}\equiv -\frac{2}{\sqrt{g}}\frac{\delta {{\mathcal{L}}_{I}}}{\delta {{g}^{\mu \nu }}}=-\frac{2}{\sqrt{g}}\frac{\partial {{\mathcal{L}}_{I}}}{\partial {{g}^{\mu \nu }}}\,\,,\,\,\,{{{\hat{T}}}_{\mu \nu }}\equiv -\frac{2}{\sqrt{{\hat{g}}}}\frac{\delta {{\mathcal{L}}_{I}}}{\delta {{{\hat{g}}}^{\mu \nu }}}=-\frac{2}{\sqrt{{\hat{g}}}}\frac{\partial {{\mathcal{L}}_{I}}}{\partial {{{\hat{g}}}^{\mu \nu }}}
\end{equation}

Since the energy momentum tensors are derived from the same interaction, we demand that the sum of variations $-\frac{\sqrt{g}}{2}{{T}_{\mu \nu }}\delta {{g}^{\mu \nu }}-\frac{\sqrt{{\hat{g}}}}{2}{{\hat{T}}_{\mu \nu }}\delta {{\hat{g}}^{\mu \nu }}$ is an exact differential and can be integrated:
\begin{equation}
\label{esix}
\frac{\partial \left( \sqrt{g}{{T}_{\mu \nu }} \right)}{\partial {{{\hat{g}}}^{\mu '\nu '}}}=\frac{\partial \left( \sqrt{{\hat{g}}}{{{\hat{T}}}_{\mu '\nu '}} \right)}{\partial {{g}^{\mu \nu }}}
\end{equation}
A variation of a determinant with respect to the matrix yields the determinant multiplied by the matrix. Therefore the metric determinants are convenient candidates for the role of the independent variables of the scalars $F,\hat{F}$, but they can not be the only ones. The only way to compose a scalar from the determinants which are scalar densities of equal weight is by their ratio. Therefore the most general scalars that depend only on the determinants are $F(S),\hat{F}(S)$ where $S\equiv \frac{g}{{\hat{g}}}$. Placing $\delta g=-g{{g}_{\mu \nu }}\delta {{g}^{\mu \nu }},\delta \hat{g}=-\hat{g}{{{\hat{g}}}_{\mu \nu }}\delta {{{\hat{g}}}^{\mu \nu }}$ in equation \eqref{esix} one obtains:
\[\sqrt{g}{{g}_{\mu \nu }}F'(S){{{\hat{g}}}_{\mu '\nu '}}=-\sqrt{{\hat{g}}}{{g}_{\mu \nu }}\hat{F}'(S){{{\hat{g}}}_{\mu '\nu '}}\]
Both metrics describe a Lorentzian space-time so that $g,\hat{g}>0$; therefore the above condition is maintained if $F'(S)$ and $\hat{F}'(S)$ have different signs for every possible metric configuration. On the other hand, the boundary conditions \eqref{efive} satisfy the demand that $F=\hat{F}=0$ when ${{g}_{\mu \nu }}={{{\hat{g}}}_{\mu \nu }}={{\eta }_{\mu \nu }}$ i.e. when $S=1$, and with the demand for non-negativity of $F,\hat{F}$ the conclusion is that $S=1$ is a minimum point of the two functions $F,\hat{F}$. An existence of a common minimum point is in contradiction to the different signs of the derivatives, therefore the statement that the determinants can not be the only variables for the scalar functions $F,\hat{F}$ is proven. 
\\
One way to overcome the problem is to try dependence on other invariants of the two metrics as variables other than the metrics, e.g. the trace. We choose another way, which is to put the determinants in the same role and to couple to other fields in order to create a scalar:
\begin{equation}
\begin{gathered}{T_{\mu\nu}}=-{\left({\sqrt{g}+\sqrt{\hat{g}}}\right)^{-4k}}\det{\left({A_{\alpha\beta}}\right)^{2k}}{g_{\mu\nu}}\hfill\\
{{\hat{T}}_{\mu\nu}}=-{\left({\sqrt{g}+\sqrt{\hat{g}}}\right)^{-4k}}\det{\left({A_{\alpha\beta}}\right)^{2k}}{{\hat{g}}_{\mu\nu}}\hfill
\end{gathered}
\label{eseven}
\end{equation}
where $k$ is an integer number, and ${{A}_{\alpha \beta }}$ is some tensor field which does not involve the metrics (no index raising). Its determinant is a scalar density with the same weight as $g,\hat{g}$. We take, for example, a tensor ${{A}_{\alpha \beta }}\equiv {{\phi }_{,\alpha }}{{\phi }_{,\beta }}$ where $\phi $ is a scalar field.\\
The energy-momentum tensors in \eqref{eseven} maintain the condition \eqref{esix} for an exact differential, and we integrate to obtain the corresponding interaction:
\begin{equation}
\int{\frac{-\sqrt{g}}{2}{{T}_{\mu \nu }}d{{g}^{\mu \nu }}=\int{-\frac{1}{2\sqrt{g}}Fdg}=\frac{{{\left( \det ({{\phi }_{,\alpha }}{{\phi }_{,\beta }}) \right)}^{2k}}}{4k-1}}{{\left( \sqrt{g}+\sqrt{{\hat{g}}} \right)}^{-4k+1}}
\end{equation}
The same result is obtained when the twin energy momentum tensor density $\sqrt{{\hat{g}}}{{{\hat{T}}}_{\mu '\nu '}}$ is integrated. One can add a mass term for the scalar field:
\begin{equation}
\label{eeight}
{{\mathcal{L}}_{I}}=\frac{{{\left( \det ({{\phi }_{,\alpha }}{{\phi }_{,\beta }}) \right)}^{2k}}}{4k-1}{{\left( \sqrt{g}+\sqrt{{\hat{g}}} \right)}^{-4k+1}}+\left( \sqrt{g}+\sqrt{{\hat{g}}} \right){{\phi }^{2}}
\end{equation}
We notice that the second addend also contributes energy momentum tensors that obeys the dominant energy condition. The interaction may also contain standard couplings of matter to the metric ${{g}_{\mu \nu }}$. For these matter fields we assume that their energy momentum tensors obey the dominant energy condition. The total energy momentum tensors for both metrics are then addends of tensors that obey the dominant energy condition, and therefore obey dominant energy condition themselves.\\
The expression in \eqref{eeight} is a source of energy momentum tensors that satisfies the dominant energy condition \eqref{gthree} and the appropriate boundary conditions \eqref{gtwo}. Based on the positive energy theorem for (multi)bimetric theories \cite{Talshir} the lowest energy is obtained when all metrics are Minkowskian and all energy momentum tensors are zero. In the case of the interaction described in \eqref{eeight}, zero energy momentum tensors are obtained only for $\phi=0$, because of the mass term. The interaction \eqref{eeight} is therefore a non-trivial bimetric interaction that constitutes, along with the Einstein-Hilbert Lagrangians with positive coefficients in \eqref{one} (with and without standard matter interaction), a bimetric model with a stable ground state.

\section{Summary and discussion}
In this paper we have shown explicitly the existence of bimetric theories for which the total canonical energy is non-negative, and is zero only for the unique vacuum state. We have used a result about the total energy of multi-metric theories, and a bimetric positive energy theorem. We have given general conditions for a stable bimetric theory, and constructed a specific model that obey these conditions, i.e. dominant energy conditions and consistent boundary limits. The model we present in this paper is a simple example of a solution to the energy stability conditions, and whether it survives other standard tests is a matter for further research. In any case, a modified model that satisfies the main conditions presented in the paper may be constructed.  \\

Bimetric theories that can be linearized usually suffer from the Boulware-Deser ghost, unless they have a specific interaction structure that sets extra constraints. The theories presented in this paper can be linearised, and although not directly proven as immune to such ghosts, they obey the dominant energy condition by definition and thus have a stable vacuum state. The numeric value of the Hamiltonian on the constraint surface is always non negative, and bounded from below by the value zero that is obtained in the unique double-Minkowski vacuum.
\\ 
The method of proving stability presented in this paper may be applied for theories that can not be linearised near the double Minkowski metric, as Milgrom's BiMOND \cite{Milgrom1}.
These theories are not threatened for now by the ghost instabilities that have been proven \cite{inconsistencies} for interacting bimetric theories that can be described in the free limit as a sum of Pauli-Fierz actions. If in the future these inconsistency theorems are extended to apply to theories that can be linearised near other background metrics, then one can use definitions for energy for general relativity with non-Minkowskian boundary conditions \cite{Abbott} and a positivity theorem for such systems \cite{Gibbons}, and attempt to generalise it for bimetric models. 

\subsection*{Acknowledgements}
I want to thank Lawrence P. Horwitz for his careful reading and useful discussions.
I gratefully acknowledge financial support from Ariel University.

\end{document}